\documentclass{eas}
\usepackage{graphicx}
\TitreGlobal{Extrasolar Planets in Multi-Body Systems, Torun 2008}

\begin{document}

\title{{Mean-Motion} Resonances of High Order in Extrasolar Planetary Systems} 
\runningtitle{{H. Rein \& J.C.B. Papaloizou: Formation of High Order {MMRs}}} 
\author{Hanno Rein}\address{Department of Applied Mathematics and Theoretical Physics; University of Cambridge\\\email{hr260@cam.ac.uk}}
\author{John C. B. Papaloizou}\sameaddress{1}

\begin{abstract}
Many multi-planet systems have been discovered in recent years. Some of them are in {mean-motion} resonances (MMR). Planet formation theory was
successful in explaining the formation of 2:1, 3:1 and other low resonances as a result of convergent migration. 
However, higher order resonances require high initial orbital eccentricities in order to be formed by this process and these are in general unexpected in a dissipative disk.
We present a way of generating large initial eccentricities using additional planets. This procedure allows us to form high order MMRs and predict new planets using a genetic {$N$}-body code. 
\end{abstract}
\maketitle

\section{High Resonances}
Of the recently discovered 322 extrasolar planets, at least 75 are in multi-planet systems (Schneider \cite{schneider}). About 10\% of these are confirmed to be in or very close to a resonant configuration. 

Two planets can easily be locked into a low order
resonance such as 2:1 or 3:2 by dissipative forces.
This process is well understood (see eg. {Lee \& Peale
\cite{leepeale}}, Papaloizou \& Szuszkiewicz \cite{papaloizou}). However, period ratios of multi-planetary
systems seem to be non-uniformly
distributed up to large period ratios of order 5.  Furthermore, there is an observational bias against
the detection of resonant systems in general (Anglada-Escude \etal{}  \cite{anglada}). One should note
that a period ratio of e.g. approximately 5 does not
imply that the system is indeed in a 5:1 {mean-motion}
resonance. We need long term, high precision measurements to
determine exactly in which dynamical state the planets are.
For most observed systems, this has not been achieved yet.
However, we hope that the increasing number of
detected planets will enable us to draw a conclusion in
a statistical manner in the near future.

These high order resonance candidates cannot be explained by
adiabatic migration only. The capture requires a high
initial eccentricity which is not expected to be present
in the system shortly after formation of the proto-planet cores.

In {Figure} 1, the final period ratio of two migrating
Jupiter mass planets is plotted as a function of the
initial eccentricity of the outer planet and the eccentricity damping
parameter $K$ which is defined as the ratio of the timescales $\tau_a$ and $\tau_e${, being the migration timescale and the eccentricity damping timescale, respectively.}   
The color encodes the final period ratio at the end of the simulation. 
At the beginning the planets are put far away from each
other. The outer planet migrates inwards on a
timescale of $\tau_a=25000$~years. One can see that a small
initial eccentricity always results in a 2:1 mean motion
resonance. At slightly higher eccentricities ($e \sim 0.05$)
the favored outcome is a 3:1 resonance. At large
eccentricities ($e \sim 0.15$) it is possible to get high
resonances like 4:1 and 5:1. Note that if $K$ is large, the eccentricities get damped before the planets {are captured} into resonance and therefore favor lower order resonances.

\begin{figure}[htb]
\begin{center}
  \includegraphics[angle=270,width=0.7\columnwidth]{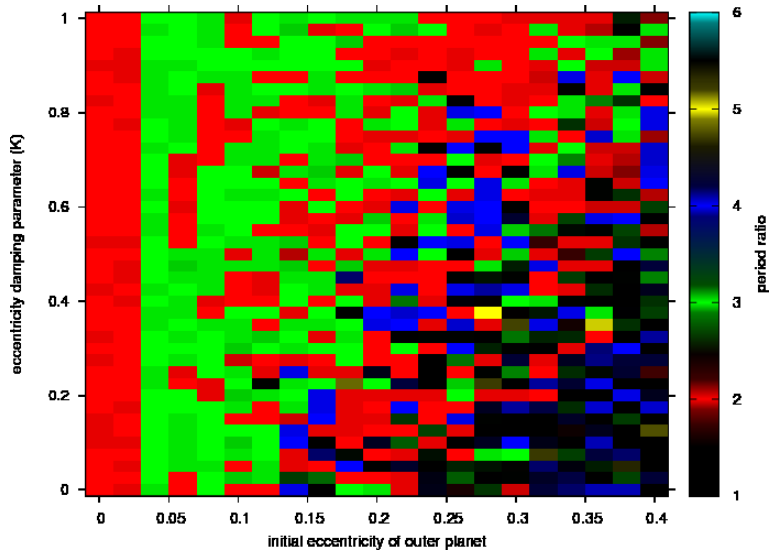}
  \caption{This figure shows the final period ratio of two convergently migrating Jupiter mass planets as a function of the initial eccentricity $e_{init}$ of the outer planet and the ratio of damping timescales $K$ (see text). 
{Red, as indicated on the color bar,
corresponds to a 2:1, green to 3:1,
blue to 4:1 and yellow to 5:1 commensurabilities
respectively. Black indicates that the period ratio
is not close to an integer value or the planets scattered within a short period of time.}}
\end{center}
\end{figure}

\newpage
\section{Additional Planets}
There are different ways to create the required initial
eccentricity for planets to get captured into high order resonances.
One possibility is a scattering event in the past. The planets could either be scattered directly into a MMR or close to it (Raymond \etal{}  \cite{raymond}).
In the latter case, migration due to a fraction of the proto-planetary disc that is still present can lead to a MMR.  
Another possibility is the presence of an additional
planet in the system. If this planet is already in a low order 
{mean-motion} resonance with one of the other planets, an equilibrium eccentricity $e_{eq}>0$ can be sustained (see e.g. Nelson \& Papaloizou \cite{nelsonpapaloizou}).

In {Figure} 2 the evolution of the semi-major axes
 and eccentricities of three planets {is} shown. The inner, middle and outer planets have masses of
1.3~$\mbox{M}_{\mbox{\scriptsize Jup}}$, 1.0~$\mbox{M}_{\mbox{\scriptsize
Jup}}$ and
0.7~$\mbox{M}_{\mbox{\scriptsize Jup}}$, respectively. Here, the masses of the inner and middle planet are similar to those in the observed system HD108874 that is close to a 4:1 MMR (see e.g. Vogt \etal{} \cite{vogt05}, Butler \etal{} \cite{butler06}).
The outer planet was put in artificially and has {not} been detected yet. However, due to its low mass and
the presence of the MMR it may actually be hiding in the system.	

All three planets feel dissipative disc forces. The migration rates differ such that the orbits of the planets converge.
Thus, the outer planet captures the
middle planet into a 2:1 resonance. They migrate together
and increase their eccentricities until they reach an equilibrium value and migrate adiabatically. Finally all three planets 
are locked into a 8:2:1 resonance. 

This solution was found using a genetic algorithm. 
It is well known that the general multi-body problem is chaotic. Formation scenarios for an observed multi-planetary system are therefore hard to find. So far, most studies have found solutions by changing simulation parameters by hand. This process can take a long time and requires physical insight. 
We use a genetic {$N$}-body
code that finds the formation scenario that fits best with the observed system. {In our code initial conditions are stored in a genom consisting of genes. The initial conditions are then integrated forward in time. After each iteration, a good genom is passed on to the next generation (iteration) of initial conditions. Each gene can also mutate with a given probability, thus creating new initial conditions. 
In the case presented above, we define a genom to be good if the resulting planetary system shows large similarities with the observed system in the eccentricities and period ratios. 
The genetic code we implemented proved to be {both} very efficient and 
stable (in terms of convergence) for the problem considered here.}

\begin{figure}[htb]
\begin{center}
  \includegraphics[angle=270,width=0.8\columnwidth]{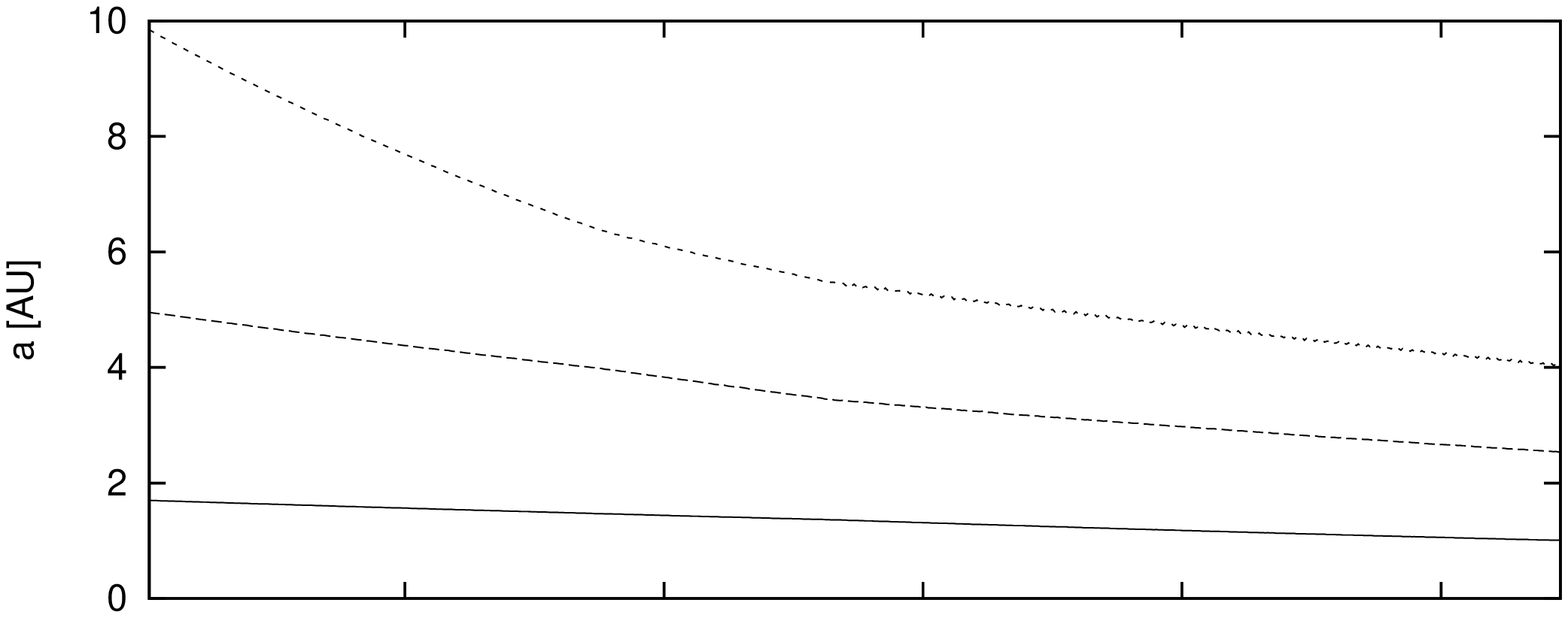}
  \includegraphics[angle=270,width=0.8\columnwidth]{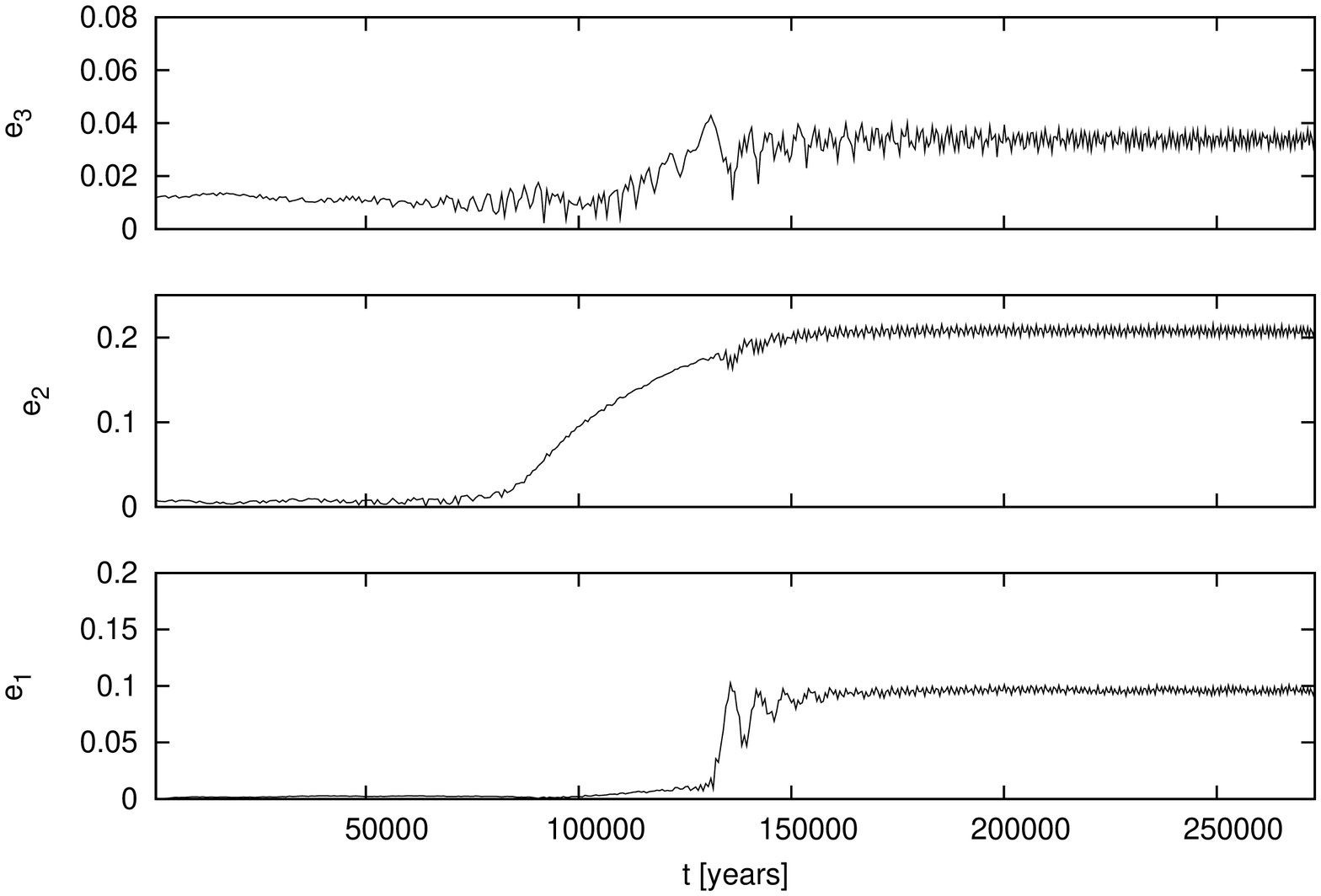}
  \caption{Top figure: evolution of the semi major axes of the outer, middle and inner planet respectively. Bottom figures: evolution of the eccentricities of the outer, middle and inner planet respectively. The planets migrate inwards convergently on slightly different time scales. The outer planet captures the middle one in  a 2:1 resonance after $\sim 10^5$ years. Their eccentricities rise, making a capture into the 8:2:1 resonance after $\sim 1.5 \cdot 10^5$ years possible. Finally all planets migrate together while the eccentricities remain constant.  }
\end{center}
\end{figure}

\section{Conclusions}
It is possible to model the observed orbital parameters of systems with high order commensurabilities in a natural way by postulating additional planets. The {solution presented here} does not rely on the random effects of {planet-planet} scattering. The prediction of
new planets can result in a test for planet formation theories and provides attractive targets for observers.

An interesting point that hasn't been discussed here is the question whether this scenario is stable to small perturbations. These perturbations could result from a turbulent disc or other planets (see e.g. Rein \& Papaloizou \cite{rein}).

Further work is underway, including hydrodynamical simulations confirming the solutions found by the {$N$}-body code.

\end{document}